\documentclass[twocolumn,preprintnumbers,amsmath,amssymb]{revtex4}

\usepackage{graphicx}
\usepackage{dcolumn}
\usepackage{bm}
\usepackage{enumerate}  
\usepackage{color}

\begin{document}


\title{Quantum interference in a macroscopic van der Waals conductor}

\author{C. W. Rischau$^1$, S. Wiedmann$^2$, G. Seyfarth$^{3,4}$, D. LeBoeuf$^{4}$, K. Behnia$^1$ and B. Fauqu\'e$^{1,5}$}%

\affiliation{
$^1$ ESPCI ParisTech, PSL Research University; CNRS; Sorbonne Universit\'es, UPMC Univ. Paris 6; LPEM, 10 rue Vauquelin, F-75231 Paris Cedex 5, France \\
$^2$ High Field Magnet Laboratory (HFML-EMFL) and Institute for Molecules and Materials, Radboud University, Toernooiveld 7, 6525 ED Nijmegen, The Netherlands \\
$^3$ Univ. Grenoble Alpes, LNCMI, F-38042 Grenoble Cedex 9, France\\
$^4$ CNRS, Laboratoire National des Champs  Magn\`etiques Intenses LNCMI (UJF, UPS, INSA), UPR 3228, F-38042 Grenoble Cedex 9, France\\
$^5$ IPCDF, Coll\`ege de France, 75005 Paris, France\\}

\date{\today}

\begin{abstract}
Quantum corrections to charge transport can give rise to an oscillatory magnetoconductance, typically observed in mesoscopic samples with a length shorter than or comparable with the phase coherence length. Here, we report the observation of magnetoconductance oscillations periodic in magnetic field with an amplitude of the order of $e^2/h$ in macroscopic samples of Highly Oriented Pyrolytic Graphite (HOPG). The observed effect emerges when all carriers are confined to their lowest Landau levels. We argue that this quantum interference phenomenon can be explained by invoking moir\'e superlattices with a discrete distribution in periodicity. According to our results, when the magnetic length $\ell_B$, the Fermi wave length $\lambda_F$ and the length scale of fluctuations in local chemical potential are comparable in a layered conductor, quantum corrections can be detected over centimetric length scales.
\end{abstract}

\pacs{Valid PACS appear here}
\maketitle


The Boltzmann equation  provides a successful description of the flow of electrons in macroscopic conductors in most cases. In this semi-classical picture, wave-like electrons (with a Fermi wave vector, $k_F$) between two scattering events separated by a typical distance, $\ell_e$ (dubbed the mean-free-path), follow the laws of classical mechanics. In this framework, the magnetoconductictivity $\sigma$ is equal to \cite{Ziman}:

\begin{equation}\label{Eq.1}
 \sigma=\frac{\sigma_0}{1+\mu^2B^2}
\end{equation}

 where $\sigma_0$=$\sigma(B=0)$ and $\mu$, the mobility, can be expressed in terms of these two length scales and the magnetic length $\ell_B$(=$\sqrt{\hbar/eB}$) as $\sigma_0$=$\frac{e^2}{h} k_{F}^{2} \ell_{e}$ and $\mu B$=$\frac{\ell_{e}}{k_{F}\ell^{2}_{B}}$. Absent from this semi-classical treatment are purely quantum effects, which give rise to non-monotonous magnetoconductivity. Landau quantization leads to an oscillating component periodic in the inverse of magnetic field \cite{Schoenberg84}. A second category of purely quantum corrections to Eq. 1 is produced by the coupling of the vector potential to the phase of the electron wave-function in real space. Two prominent examples  are Universal Conductance Fluctuations (UCF) and the Aharanov-Bohm (AB) effect, which lead to a non-monotonous magneto-conductivity periodic in magnetic field with an amplitude of the order of the quantum of conductance $e^2/h$ \cite{Ihn}. They emerge whenever the amplitude of the potential vector fluctuates over a length scale shorter than the electron phase coherence length, $\ell_{\phi}$, which is usually in the range of microns. The fate of these quantum corrections when the distance between electrodes exceeds $\ell_{\phi}$ by far is an open question. In this paper, we report on the observation of a purely quantum correction to magnetoconductivity in centrimetric samples of Highly Oriented Pyrolytic Graphite (HOPG) samples. Oscillations of magnetocondutance, periodic in magnetic field were detected below 2 K , when the magnetic field was strong enough to make $l_B$ comparable to the Fermi wave length $\lambda_F$. We propose a moir\'e superlattice with a periodicity, $D$, as large as 50 nm, as the origin of the observed quantum interference. Our result shows that it is indeed possible to detect quantum interference phenomena in a  three dimensional macroscopic matrix in the limit where $l_B<\lambda_F<D$.\\
\begin{figure}[h]
\begin{center}
\includegraphics[angle=0,width=8.5cm]{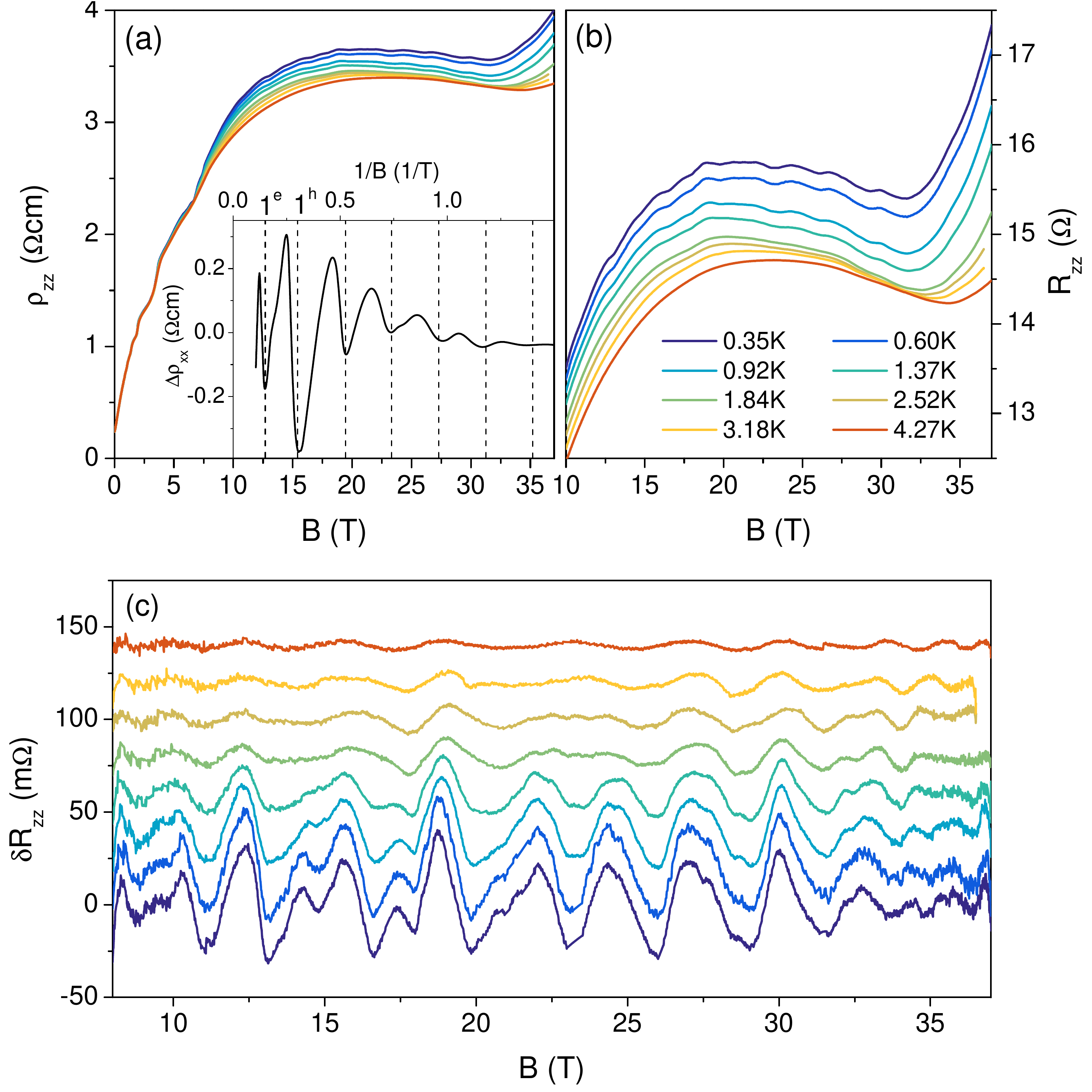}
\caption{a) Longitudinal $c$-axis magnetoresistivity $\rho_{zz}$ for a HOPG sample (sample E) as a function of magnetic field $B$ up to 37 T. The inset shows the low field quantum oscillations as a function of $B^{-1}$. Above 7.5 T, both holes and electrons are confined in the lowest Landau levels ($n=0$). b) Longitudinal resistance $R_{zz}$ in the magnetic field range of 10-30 T. Above 10 T, unexpected additional resistance oscillations are observed, which become more visible after a smooth background subtraction as shown in panel c). While the low-field quantum oscillations are only weakly temperature dependent, these oscillations quickly fade out with temperature.}
\label{Fig1}
\end{center}
\end{figure}



A magnetic field of only a few Tesla is large enough to confine all the carriers in graphite to their lowest Landau levels, a situation called the quantum limit. As illustrated in the insert of Fig. \ref{Fig1}, up to 7.5 T, the longitudinal magnetoresistance ($\rho_{zz}$)  shows oscillations periodic in the inverse of the magnetic field, as a result of successive passage of Landau levels through the Fermi level. The Fermi surface of graphite is formed by one electron and one hole pocket, elongated along the (H-K-H) and (H'-K'-H') valleys of the hexagonal Brillouin zone \cite{SW1958,McClure1957} and has been confirmed in numerous studies of quantum oscillations \cite{Schneider2009,Hubbard2011}. A sketch of the Fermi surface and the Brillouin zone is reported in the SM. Above 7.5 T, all the carriers are confined in the $n=0$ spin-split Landau level. In the simplest case, besides the depopulation of the (0,+) LLs  expected to occur at 37 T \cite{Takada1999}, no field scale should be present. Yet, as can be seen in Fig. \ref{Fig1} a), several additional features can be observed in $\rho_{zz}$.\\

\begin{figure}[htbp]
\begin{center}
\includegraphics[angle=0,width=8.5cm]{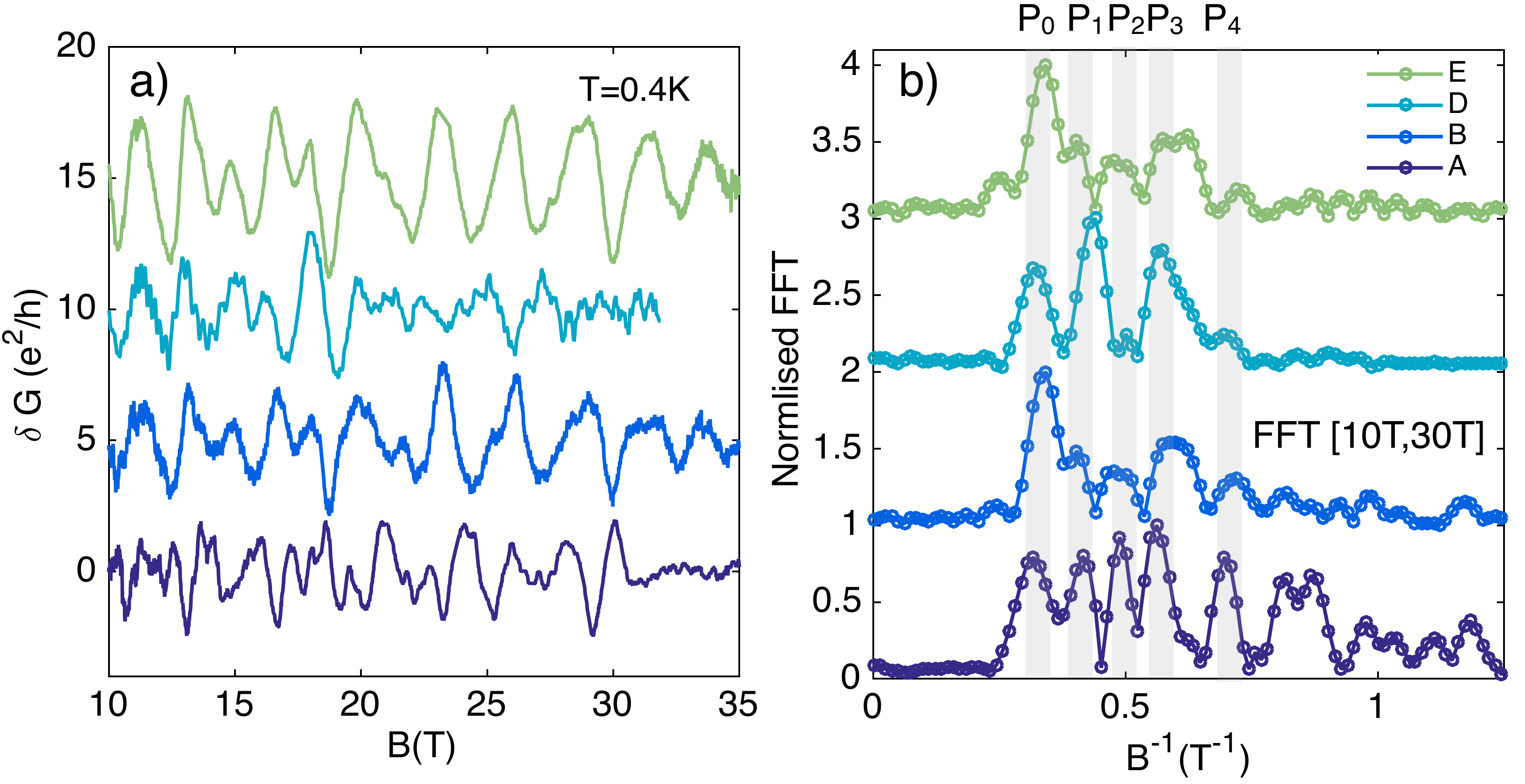}
\caption{ a) Oscillating part $\delta G$ of the conductance in units of $e^2/h$ as a function of magnetic field measured on different HOPG samples at $T=0.4$ K in both transverse (sample A) and longitudinal (samples B, D and E) configuration. The curves are shifted for clarity. b) Fast Fourier Transform (FFT) spectra of $\delta G$ for the same samples. Despite differences in the oscillation patterns, all spectra display the same characteristic frequencies that only differ in their FFT amplitude from one sample to another.}
\label{Fig2}
\end{center}
\end{figure}

First, above 30 T, we observe an increase of $\rho_{zz}$, which shifts to higher magnetic fields with increasing temperature. This increase is the onset of a field-induced state discovered in the 1980s (see \cite{Yaguchi2009} for a review). Although there is currently no consensus on the nature of this phase, it has been attributed to the formation of a density wave state in the Landau levels (0,$\pm$) mediated by the electron-electron interactions. Recent experiments extended up to 80 T have revealed that two successive field-induced instabilities occur \cite{Fauque2013} (instead of just one as believed previously \cite{Yaguchi2009}). The second feature visible in  Fig. \ref{Fig1}, and the main subject of this paper, is the emergence of  additional anomalies between 10 and 30 T. Subtracting a smooth background (labelled $R_{zz}^{(0)}(B)$), we resolve an oscillating contribution $\delta R_{zz}=R_{zz}-R_{zz}^{(0)}$ to the total magnetoresistance, which becomes more pronounced with decreasing temperature (see Fig. \ref{Fig1} c)).\\
The robustness of these oscillations is illustrated in Fig. \ref{Fig2} a), which plots the oscillating part of the conductance $\delta G_{xx,zz}=1/R_{xx,zz}-1/R_{xx,zz}^{(0)}$ in units of $e^2/h$ for four different HOPG samples of different grades and different directions of the electrical current ($R_{xx}$ and $R_{zz}$ correspond to the transverse and longitudinal resistance, respectively). We note that for graphite, where $\rho_{xy}<<\rho_{xx}$, $G_{xx,zz}\approx R^{-1}_{xx,zz}$. While the oscillation pattern slightly changes from one sample to another, all samples show oscillations of the magnetoconductance with an amplitude of the order of $e^2/h$. Fig. \ref{Fig2} b) plots the FFT (Fast Fourier Transform) spectra of $\delta G$ for the same samples. All spectra exhibit characteristic periods (labelled $P_{0,...,4}$) of about $1-3$ T. From the FFT spectra we find that in the case of the transverse geometry ($R_{xx}$) all periods contribute with roughly the same amplitude, while they differ from one sample to another in the case of the longitudinal ($R_{zz}$) geometry.\\
While quantum oscillations are periodic in the inverse of the magnetic field, these oscillations are periodic in magnetic field, a direct signature of quantum interference in real space. 
If we extract the characteristic area of these four periods using $P_i \pi r^2_i=h/e$  as in an AB ring geometry, we find a typical radius of the order of 25 nm (see SM section C for the radius values of each periods). We assume here that $\frac{h}{e}$ oscillations prevail.These lengths are longer than the two length scales already introduced above, i.e., $\ell_B(B=10\textnormal{T})=8$ nm and the in-plane Fermi wave length $\lambda_F$($\approx14$ nm).\\

\begin{figure}[htbp]
\begin{center}
\includegraphics[angle=0,width=8.5cm]{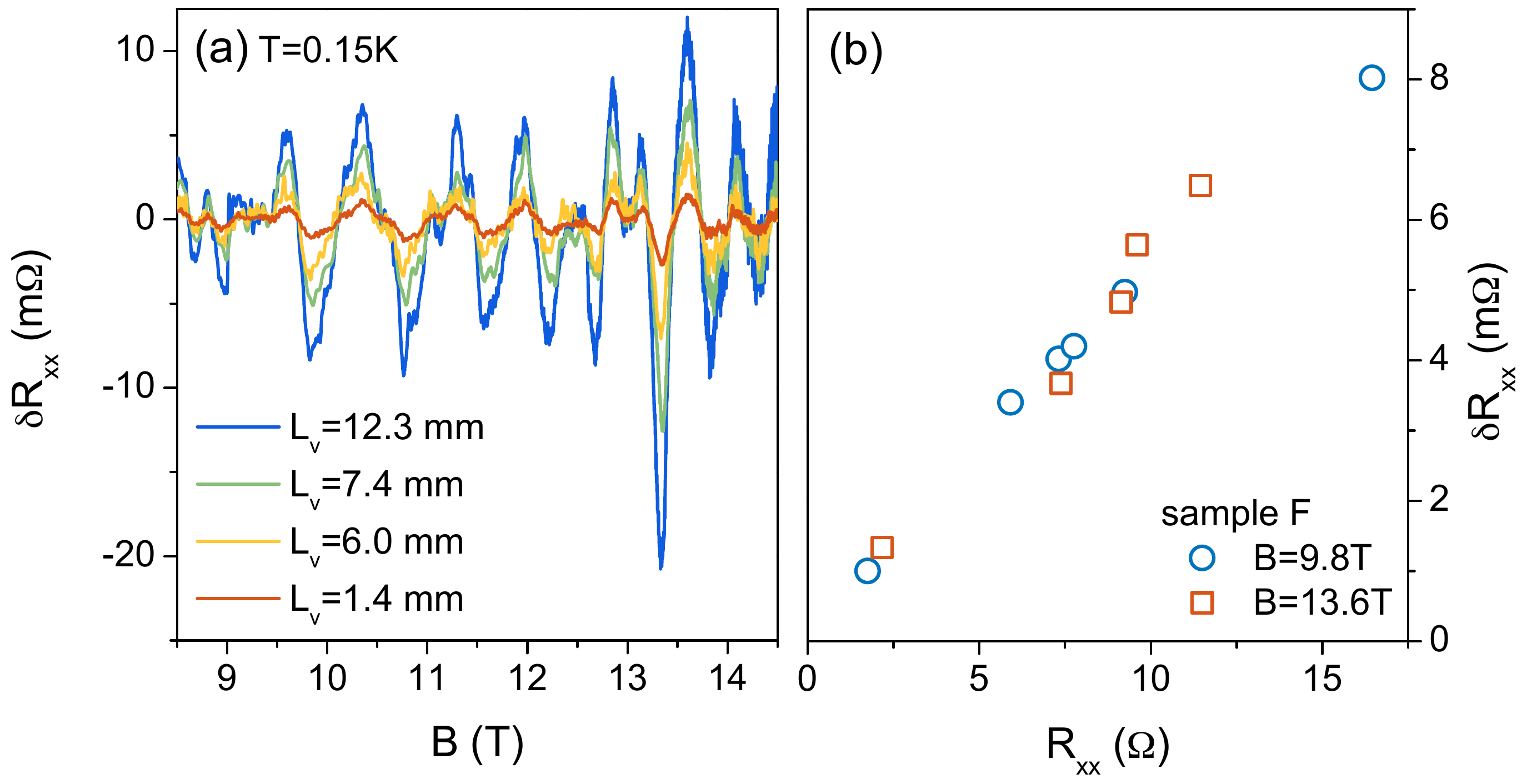}
\caption{Size dependence of the magnetoresistance oscillations ($\delta R_{xx}$): a) Comparison of the resistance oscillations $\delta R_{xx}$ measured on a sample
with a voltage lead contact length ranging from 1.4 to 12.3 mm at $T=150$ mK. b) Oscillation amplitude $\delta R_{xx}$ as a function of the resistance ($R_{xx}$) at two different magnetic fields ($B=9.8$ and 13.6 T) deduced from a).}
\label{Fig3}
\end{center}
\end{figure}

The oscillations are robust to thermal cycling, insensitive to the polarity of the magnetic field and independent of the position of the voltage or current contact leads (see Supplemental Material, Section C). However, they scale with the distance between voltage leads, labelled $L_v$. Fig. \ref{Fig3} a) presents the trace of the oscillations measured on the same sample for different $L_v$ showing that the amplitude of the oscillations increases as $L_v$ increases. The comparison between the resistance at a given field and the amplitude of the oscillations reported in Fig. \ref{Fig3} b) shows a linear correlation, demonstrating that both quantities follow Ohm's law. A similar conclusion can be drawn from the study of the thickness dependence of the oscillation amplitude reported on the SM section C. Therefore, the amplitude of these oscillations is not universal.\\

Further insights on the electronic states responsible for these oscillations can be obtained by looking at the temperature dependence of the oscillation peaks in and out of the field-induced state, as illustrated in Figs. \ref{Fig3} c) and d). For magnetic fields below 30 T (i.e., out of the field-induced state), the amplitude of the peaks continuously grows below 2 K and saturates at 350 mK. In mesoscopic physics, the temperature onset of the oscillations is generally attributed to the competition between the thermal and the Thouless  energy (the energy scale on which electrons remain coherent over the sample). In other words, it corresponds to the cut-off of $\ell_{\phi}$ by the thermal length $L_T=\sqrt{hD_c/k_BT}$ where $D_c$ is the diffusion constant (see Supplemental Material, Section B for an estimation of $D_c$). With a temperature onset of 2 K, we estimate $\ell_{\phi}\approx 1$ $\mu$m. The temperature dependence of the oscillations is different in the field-induced state at $B=34$ T: the amplitude of the oscillations only increases from 4 to 2 K, but decreases below 2 K.  We note that the amplitude of the oscillations also collapses in the field-induced state in the longitudinal configuration (see sample A in Fig. \ref{Fig2}a)). At $B=34$ T, the temperature onset of the field-induced state is estimated to lie slightly above 4 K (see Fig. \ref{Fig1} b)). It has been recently found that a gap opens up along the $c$-axis when this electronic state is induced by magnetic field \cite{Fauque2013}. Therefore, the reduction in the amplitude of the oscillations observed below 34 T suggests that the quantum interference phenomenon observed here is also destroyed by the opening of the gap in  the electronic spectrum. \\

\begin{figure}[htbp]
\begin{center}
\includegraphics[angle=0,width=8.5cm]{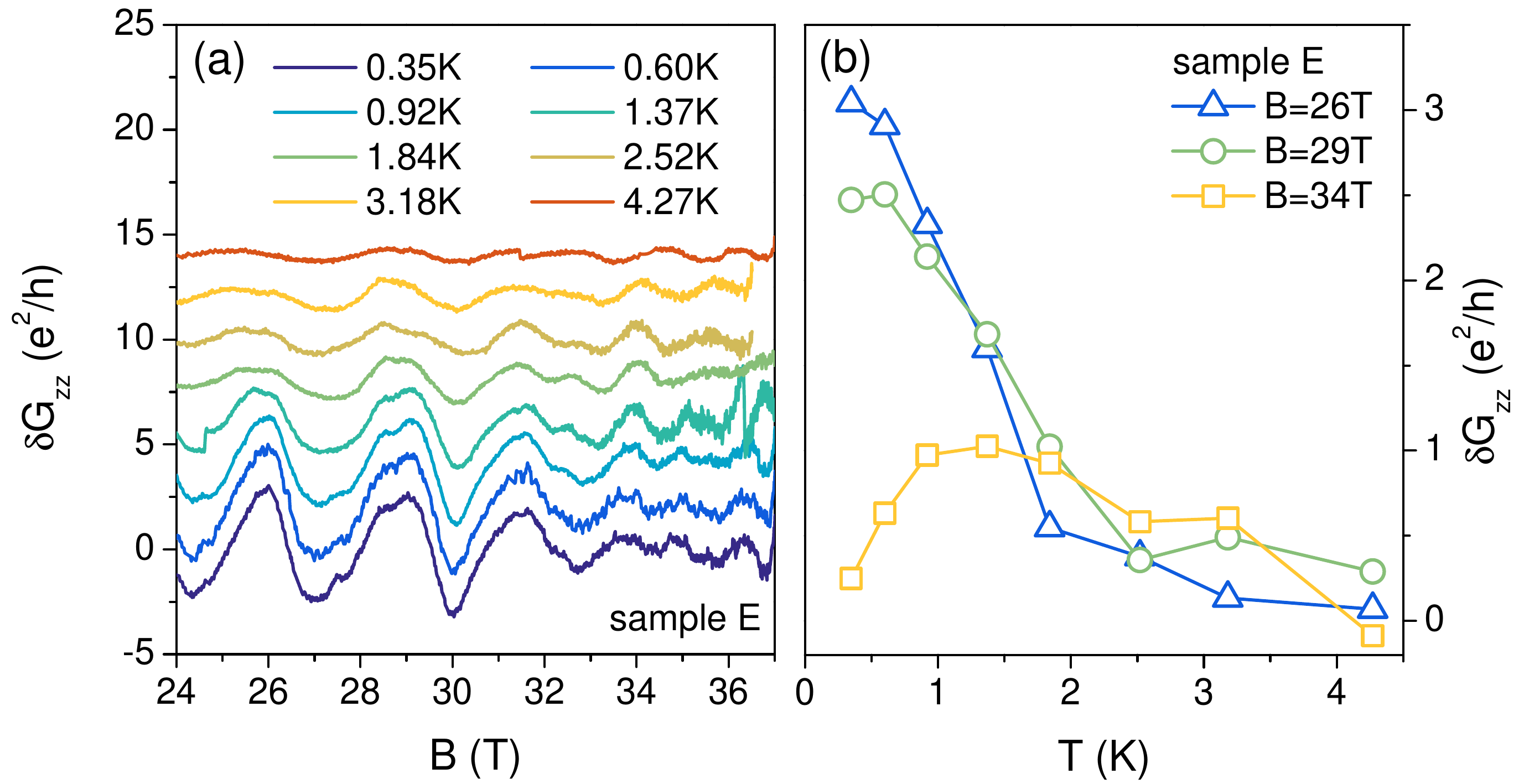}
\caption{ Temperature dependence of the magnetoconducatance oscillations a) $\delta G_{zz}$ as a function of magnetic field for temperatures from 0.35 to 4.3 K. b) Temperature dependence of $\delta G_{zz}$ at $B=26$, 29 and 34 T. In the field-induced state ($B=34$ T) we observe that the amplitude of the peak does not change considerably, while it strongly increases with decreasing temperature below the field-induced state ($B=26$ and 29 T).}
\label{Fig4}
\end{center}
\end{figure}


Our observations put interesting perspectives on recent unexplained observations in low-density conductors. 
First, in the case of graphite, recent oscillatory phenomena, such as magnetoresistance oscillations with a period of 0.8 T below 9 T and additional peaks in $\rho_{xx}$ above 10 T (concomitant with plateau-like structures in $\rho_{xy}$) \cite{Kopelevitch09b,Kopelevitch09}, have been reported. In the framework of our study, they can now be attributed to the AB-type oscillations discussed above. As reported in Fig. \ref{Fig3} a), we also find that the low field part of $\delta R_{xx}$ is dominated by small periods formed by a linear combination of the main periods P$_{0},..,P_{4}$. Based on the temperature dependence reported in Fig. \ref{Fig3} c), we attribute the high magnetic field peaks (observed at $T=1.5$ K and interpreted as a fingerprint of a fractional quantum Hall effect in Refs. \cite{Kopelevitch09b,Kopelevitch09}) to the precursor of the AB-type oscillations seen in this work. Secondly, AB resistance magneto-oscillations with similar periods have been reported in thin natural graphite samples after swift-heavy ion irradiation \cite{Latyshev10} and in samples containing a single nano hole \cite{Latyshev14}. We show that they are indeed an intrinsic property of HOPG samples. Thirdly, our results are reminiscent of the magnetoconductance oscillations observed in non-metallic samples of the doped topological insulator Ca$_x$Bi$_{2-x}$Se$_3$ \cite{Checkelsky09} exhibiting the same oscillation amplitude and period. Interestingly, both systems share a layered structure with the same bulk carrier density ($n=5\times10^{18}$ cm$^{-3}$) and the same in-plane magnetoresistance ($\rho_{xx}(20\textnormal{T})\approx 20$ m$\Omega$cm). We thus conclude that such magnetoconductance oscillations are a universal property of low-carrier density van der Waals conductors in the regime of the quantum limit.\\


The main result of this study is the observation of AB oscillations in macroscopic graphite samples in absence of any intentional attempt to introduce an array of rings. Two questions emerge from these unexpected finding. First, what it is the nature of the defect which plays the role of the ring in the classic AB geometry? Second, what it is the origin of the discrete size distribution of the defects as deduced from our FFT spectrum? 

In order to answer the first question, it is helpful to compare our result with the case of a two-dimensional electron gas (2DEG). In the quantum Hall effect regime, AB-type oscillations are suppressed due to the absence of backscattering \cite{Timp89}. However, resistance fluctuations or oscillations have been observed in the quantum Hall regime for nanowires \cite{Timp87,Main89,Simmons91} or in antidot arrays \cite{Nihey93} over macroscopic length scale \cite{Kato08}. According to the Jain-Kivelson theory \cite{Jain88}, these fluctuations or oscillations arise because of a tunneling process between opposite edge states mediated by bound states encircling a defect potential. These bound states form as a result of the Bohr-Sommerfeld quantization if an integral number of flux quanta $\frac{h}{e}$ penetrate the defect potential area. In the case of layered low density conductors such as graphite or Bi$_2$Se$_3$, we can think of two types of long-range potential fluctuations where the flux can, as well, be quantized. \\
 

In the presence of a local defect such as vacancy or anti-site defect electrons tend to screen the electrical field on the Thomas-Fermi length scale set by the Fermi wave length ($\lambda_F$). In the case of low carrier systems where $\lambda_F\approx10$ nm, the screening occurs on a length scale much larger than the actual size of the defects. In the case of doped Bi$_2$Se$_3$, charged inhomogeneities (or puddles) with a typical length scale of 20 nm have been observed in scanning tunneling microscopy (STM) \cite{Beidenkopf11}. Also in the case of graphene, large potential fluctuations have been seen in STM measurements, but were attributed to the effect of the substrate \cite{Andrei12}. The puddle scenario is appealing, but can hardly answer to the second question : in the simplest picture, one expects a continuous distribution of length scales of the puddles, and thus any oscillating pattern would smear out. \\

Alternatively we can think of another kind of lattice defect which forms at the boundary between two crystalline regions with different orientations. At this interface, where two hexagonal lattices overlap and are misoriented by an angle $\theta$, a superlattice or so-called moir\'e pattern is formed. The periodicity ($D$) of the super hexagonal lattice is given by:

\begin{equation}
\label{Dmoire}
D=\frac{d}{2 \sin(\theta/2)}
\end{equation}

where $d=\sqrt{3}a_l$ and a$_l=1.42$ $\AA$ the distance between carbon atoms. These superlattices have been studied for more than three decades in HOPG graphite  \cite{Pong05} and have been also recently observed in Bi$_2$Se$_3$ \cite{Liu10}. The typical periodicity observed in STM measurements is of the order of a few nm, but periodicities up to 44 nm, close to the diameter length scale deduced from our measurement, have also been reported for HOPG samples \cite{Oden91}. In addition the coherence length deduced from the temperature dependence of the oscillations is in good agreement with this scenario: according to STM measurements \cite{Pong05} the moir\'e pattern can span over several microns.

For a periodicity of $D=50$ nm, the angle of misorientation is as low as $\theta = 0.3^{\circ}$. According to recent experimental and theoretical works, when the misalignment angle is as small, the optimal structures differ significantly from those expected for large misalignment angles \cite{Woods14,Wijk15} where the strain induced by the moir\'e lattice is equally distributed in the layer. For low angles ($\theta<$1$^{\circ}$), when the two lattices are commensurate, the strain is concentrated on \textit{hot spots} separated by flat regions with no strain \cite{Wijk15}. In order to answer to the second question, i.e. the origin of the discrete size distribution of the defects, we propose that Eq. (2) breaks down in the commensurate phase. In this regime, when $\theta$ increases, the strain is accumulated on hot spots and the periodicity stays constant up to a threshold where it becomes more favorable to the system to increase the density of  hot spots. As a result the periodicity abruptly decreases and stays constant up to the next threshold. The periodicity follows a stair-like distribution as a function of $\theta$. For larger angles (in the incommensurate phase), the continuum of periodicity cancels out the oscillation pattern, which explains why only small periodicities are observed.

Finally, we comment on the amplitude of the oscillations. With a coherence length of about 1 $\mu$m and a periodicity of about 50 nm, the number of loops in the phase coherent region can be as high as 20. For such a large number of loops, the amplitude of the AB oscillations for a 2D metallic network (in the limit where $\ell_B>>\lambda_F$) is reduced by decoherence processes and is of the order of $10^{-2}$ $e^2/h$ \cite{Umbach86,Schopfer07}, which is two orders of magnitude smaller than what we observe. However, our result differs from the case of the 2D metallic network in at least two points. First, the oscillations are observed in the quantum limit regime where $\ell_B<\lambda_F$. Interestingly, in the case of a 2DEG it has been demonstrated that in this limit, the phase coherence plays no significant role in the pattern of the resistance fluctuations \cite{Machida89}. Second, we show that the electronic degree of freedom along the direction of the magnetic field (i.e., the third dimension) is of crucial importance. Alternatively, the system can be described as a network of independent resistances connected in series and in parallel. If each sub-unit has the same oscillating pattern of amplitude  $\delta g_0 \approx e^2/h$, the total amplitude of the oscillations ($\Delta G$) will be equal to $\Delta G =M^{\frac{1}{2}}N^{-\frac{3}{2}}\delta g_0$ where M and N are the number of sub-systems in series and parallel respectively \cite{Lee87}. In the case where $M\approx N^3$, we can resolve magnetoconductance oscillations of an amplitude $e^2/h$ over a macroscopic distance. Further work using microscopic probes should help clarify the density of the sub-units and quantify the value of M and N in HOPG samples.

 In conclusion, we find experimental evidence for quantum interference phenomenon detected in macroscopic samples of HOPG graphite when all the carriers are confined in the lowest Landau levels.  We argue that they can be explained by invoking  the presence of moir\'e superlattices with large periodicity ($D\approx 50$ nm) characterized by a discrete. Interestingly, these oscillations that have also been observed in another Van der Waals-system (Bi$_2$Se$_3$), appeared in the limit where  $l_B<\lambda_F<D$, the same limit where Hofstadter's butterfly spectrum has been recently reported in bilayer graphene \cite{Dean13, Ponomarenko13}. These results offer a new avenue to explore quantum interference process in the quantum limit of 3D conductors which, to our knowledge, have never been explored theoretically.

This work is supported by the Agence Nationale de Recherche as a part of QUANTUMLIMIT project, by a grant attributed by the Ile de France regional council. We acknowledge support from the HFML-RU/FOM and LNCMI-CNRS, which are both members of the European Magnetic Field Laboratory (EMFL). We thank B. Altshuler , B. Doucot, Y. Kopelevich, M. Goerbig, M. Katsnelson, G. Montambaux and S. Yuan for stimulating discussions.







\end{document}